\documentclass{sig-alternate}

\usepackage{multirow}
\usepackage{graphicx}
\usepackage{subfigure}
\usepackage{epstopdf}
\usepackage{color}
\usepackage{amsmath}
\usepackage{url}
\usepackage{comment}
\usepackage{etoolbox}
\usepackage{stfloats}
\usepackage{enumerate}

\newbool{inccomment}
\setbool{inccomment}{true}
\newcommand{\hl}[1]{\ifbool{inccomment}{{\color{blue}#1}}{}}
\newcommand{\cc}[1]{\ifbool{inccomment}{{\color{magenta}#1}}{}}

\makeatletter
\patchcmd{\maketitle}{\@copyrightspace}{}{}{}
\makeatother

\makeatletter
\def\@maketitle{\newpage
 \null
 \setbox\@acmtitlebox\vbox{%
\baselineskip 20pt
\vskip 1em                   
   \begin{center}
    {\ttlfnt \@title\par}       
    \vskip -0.5em                
    {\subttlfnt \the\subtitletext\par}\vskip 1.25em
    {\baselineskip 16pt\aufnt   
     \lineskip .5em             
     \begin{tabular}[t]{c}\@author
     \end{tabular}\par}
    \vskip 1.5em               
   \end{center}}
 \dimen0=\ht\@acmtitlebox
 \unvbox\@acmtitlebox
 \ifdim\dimen0<0.0pt\relax\vskip-\dimen0\fi}
\makeatother

\begin{document}

\title{A Memristor based Unsupervised Neuromorphic System Towards Fast and Energy-Efficient GAN}


\author{
    Fuqiang Liu, Chenchen Liu and Fukun Bi\\
    \email{fqliu92@gmail.com\ chchenliu@gmail.com\ bifukun@ncut.edu.cn}
    }

\maketitle




\begin{abstract}
Deep Learning has gained immense success in pushing today's artificial intelligence forward. To solve the challenge of limited labeled data in the supervised learning world, unsupervised learning has been proposed years ago while low accuracy hinters its realistic applications. Generative adversarial network (GAN) emerges as an unsupervised learning approach with promising accuracy and are under extensively study. However, the execution of GAN is extremely memory and computation intensive and results in ultra-low speed and high-power consumption. In this work, we proposed a holistic solution for fast and energy-efficient GAN computation through a memristor-based neuromorphic system. First, we exploited a hardware and software co-design approach to map the computation blocks in GAN efficiently. We also proposed an efficient data flow for optimal parallelism training and testing, depending on the computation correlations between different computing blocks. To compute the unique and complex loss of GAN, we developed a diff-block with optimized accuracy and performance. The experiment results on big data show that our design achieves $2.8\times$ speedup and $6.1\times$ energy-saving compared with the traditional GPU accelerator, as well as $5.5\times$ speedup and $1.4\times$ energy-saving compared with the previous FPGA-based accelerator.
\end{abstract}
\section{INTRODUCTION}
Deep learning has achieved great success in various artificial intelligence applications such as object detection~\cite{FastRCNN,Resnet}, tracking~\cite{FCN} and natural language processing~\cite{DNLP}. 
Normally, supervised learning is employed in the state-of-the-art applications, where a deep neural network is trained from labeled training data and desired outputs are obtained after going through an inferring function like backpropagation~\cite{CNN1,CNN2}.
The supervised learning has been proved powerful, however, challenges emerged with the fast growth of the applications complexity: large-scale of labeled data is in demand and its learning capability is constrained.

Unsupervised learning, which has the capability to learn from the unlabeled data directly appears as a possible solution. 
Nonetheless, accuracy is usually low in the conventional unsupervised learning and its feasibility in the realistic applications is hindered~\cite{unsupervised}. 
Recently, generative adversarial networks (GAN) was proposed as a promising solution for the above challenges~\cite{gan}.
GAN can estimate generative models from an adversarial process by training two models--generative and discriminate model simultaneously~\cite{gan}. 
Feature representations can be learned from unlabeled data and improved accuracy of unsupervised learning is achieved on GAN~\cite{DCGAN}.
However, the challenge of high demand for computing resource that exists in deep neural network becomes more severe in GAN computation. 
Both training and testing executions are particularly slow and energy hungry~\cite{gan,DCGAN}.

Extensively research efforts have been devoted to accelerate and improve computing efficiency of deep learning, such GPU~\cite{CNN2}, FPGA~\cite{7760779}, emerging non-von Neumann accelerators~\cite{PRIME,7920854}.
Very recently, GPU~\cite{DCGAN} and FPGA~\cite{fpgagan} based GAN computations have been proposed with significantly improved speed and energy efficiency compared with CPU.
However, the computing efficiency is still constrained as data bandwidth in these architectures is still limited and the performance improvement mainly relies on reckless resource accumulation.
Recent studies in developing novel non-von Neumann accelerators such as in-memory accelerator~\cite{7284059,article11} and neuromorphic computing system~\cite{7167197,7920854,Liu2017}, esp. the designs with novel nano-devices like memristor paved a way towards computational efficient GAN.

In~\cite{PRIME,7920854}, convolutional neural networks (CNNs) were deployed on memristor crossbar and an on-line training process with backpropagation were implemented via a hardware and software co-design methodology. 
However, these previous approaches cannot be used directly in GAN computation because of the following reasons.
First, different with traditional CNNs training, two learning models are executed simultaneously in GAN's training phase and requires an adaptive data flow for optimal computing efficiency. 
Moreover, backpropogation is simplified in a hardware-friendly way that the initial error in training is considered as the difference between the true label and the predicated label, which works conditionally under the prediction that the cost function of CNN is cross entropy.  
This assumption is invalid in GAN, therefore, the memristor-based hardware implementation in previous works cannot be utilized.

In this work, we developed a memristor-based unsupervised neuromorphic system for a fast and energy-efficient GAN computation to solve the above challenges. Our contributions can be summarized as follows:

\begin{itemize}
\item We exploited a hardware and software co-design approach to map the computation blocks in GAN to the memristor-based crossbars efficiently.
The computing system is composed of three major blocks--\emph{Generator}, \emph{Discriminator}, and \emph{Diff} block.
The \emph{Diff} block is designed  to compute the cost function of GAN accurately with low hardware cost;
\item We proposed an adaptive data flow for GAN with optimal computation parallelism and efficiency. In forward phase, the \emph{Generator} and \emph{Discriminator} block worked in parallel to generate artificial data and extract features; In backward phase, the  \emph{Generator}  and \emph{Discriminator} block were trained effectively with the initial errors computed by the {Diff} block. 


\item We evaluated the system accuracy in different data precision and the system performance in speed and energy. The proposed system performance on ImageNet and Lsun/bedroom was also compared with the GPU-based and FPGA-based GAN computation in previous works.
\end{itemize}

The experimental results show that our proposed design can achieve $2.8\times$ ($2.7\times$) and $5.5\times$ ($4.8\times$) speedup, as well as $6.1\times$ ($6.1\times$) and $1.4\times$ ($1.5\times$) energy-saving compared with GPU-based~\cite{DCGAN} and FPGA-based~\cite{fpgagan} GAN accelerators respectively on Lsun (ImageNet) dataset.

\section{BACKGROUND}
\subsection{Generative Adversarial Networks}\label{sec:GAN}

The generative adversarial network was developed as an unsupervised model that can learn effective feature representations from unlabeled data while having improved accuracy compared with traditional unsupervised learning~\cite{gan,DCGAN}.
Two learning models form the GAN: a generator and a discriminator. 
Normally, the generator is a deconvolutional neural network (Deconv-NN) for artificial data generation, and 
the discriminator is a convolutional neural network (CNN) for distinguishing the artificial data from real data. 
The training phase of GAN involves two major learning procedure--effective generator and discriminator learning based on backpropagation.
The target of the training process is obtaining a generator that can generate most likely the same as the true CNN training data and a discriminator that can extract feature effectively. 
The training process is can be summarized as four major procedures, which are named as $D_{forward}$, $D_{back}$, $G_{forward}$, and $G_{back}$ in this work.

\begin{itemize}
\item $D_{forward}$ computes the cost function to obtain the error that should be transmitted to the discriminator for its backward weight updating.
More specifically, with a batch of $m$ noise samples--${z^{1},...,z^m}$ are given as inputs, the generator generates $m$ artificial samples and this process is defined as $G(z^{i})$. 
The CNN based discriminator processes the $m$ artificial samples (i.e. $G(z^{i})$) and $m$ real samples (i.e. ${x^{1},...,x^m} $) through forward computations, and then cost function is computed following Equation~\ref{equ:cost function1}.

\begin{equation}
\label{equ:cost function1}
Error_{D}=\triangledown _{\theta_{d}} \frac{1}{m}\sum_{i=1}^{m}[logD(x^i)+log(1-D(G(z^i)))] 
\end{equation}

\item $D_{back}$ updates the weights of the discriminator by ascending the stochastic gradient obtained from Equation~\ref{equ:cost function1}, i.e. $Error_{D}$.

\item $G_{forward}$ computes the cost function to gain the error that should be given to the generator for its weight updating. 
The cost function is computed as Equation~\ref{equ:cost function2}. 

\begin{equation}
\label{equ:cost function2}
Error_{G} = \triangledown _{\theta_{g}} \frac{1}{m}\sum_{i=1}^{m}log(1-D(G(z^i)))
\end{equation}

\item $G_{back}$ updates the weights of the generator by ascending the its stochastic gradient obtained from Equation~\ref{equ:cost function2}, i.e. $Error_{G}$.
\end{itemize} 

Here, $G(\cdot)$ and $D(\cdot)$ represents the generator and discriminator respectively, $x$ is the real data, and $z$ is the noise given to the generator.

\subsection{Memristor Crossbar for (De)Convolutional Computation }\label{sec:crossbar}

The limited data bandwidth, as well as the performance gap between processing units and memory of the conventional computing platform becomes a major obstacle in the deep leaning based applications.
Novel computing platforms, such as the in-memory accelerator~\cite{7284059}, neuromorphic computing~\cite{7464347}, etc. therefore have been extensively investigated as a promising solution. 
The emerging of novel nano-devices such as spin, phase change and memristor device  also accelerates its development  and corresponding accelerators are developed accordingly. 
Among them, the memristor based computing platform attracts people's attention own to the high density, high speed, multiple level states, etc~\cite{5430304,PRIME,7920854}. 
In this work, the memristor-based computation platform for GAN is developed.
The generator that based on deconvolutional network and the discriminator based on convolution network of GAN is deployed on the memristor crossbar structure.



\begin{figure}[!t]
\centering 
\subfigure[]
{\label{fig:crossbar:a} \includegraphics[width=0.4\textwidth]{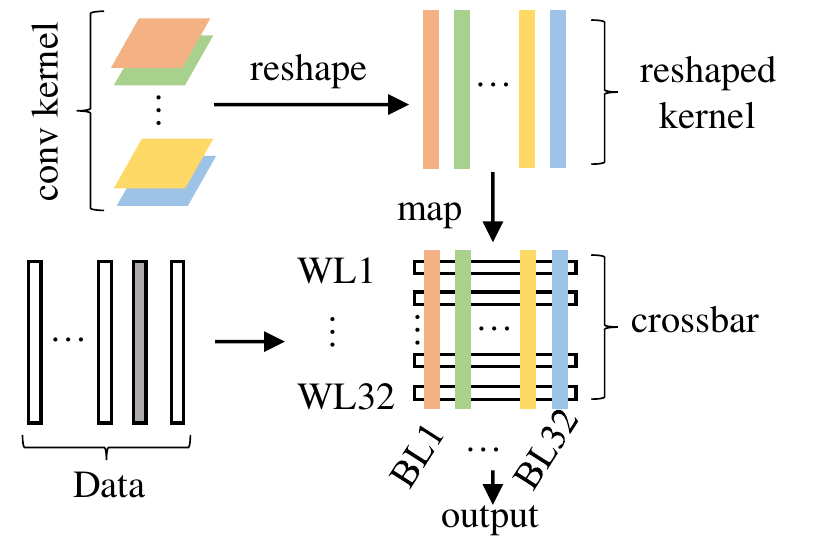}} \\
\subfigure[]
{\label{fig:crossbar:b} \includegraphics[width=0.4\textwidth]{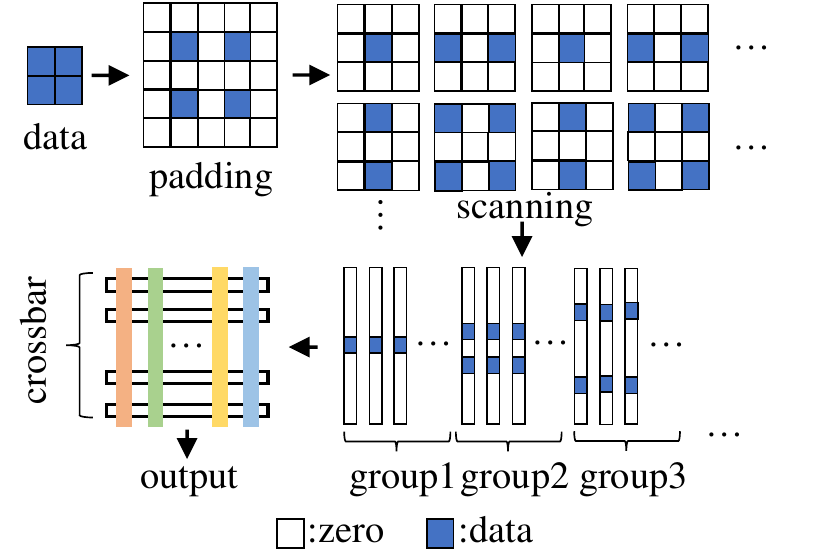}} 
\caption{Implementation of CNN (a) and Deconv-NN (b) on memristor crossbar} 
\label{pic:cnn_crossbar} 
\end{figure}

In previous work, convolutional networks have been deployed on the memristor-based crossbar structure, as is depicted in  Figure~\ref{pic:cnn_crossbar} (a)~\cite{7920854}. 
For example, to deploy a convolutional layer with 32 kernels, each kernel is reshaped to a vector that can be programmed to a memristor crossbar.
And the inputs data is given to the memristor crossbar to execute the dot-productions computations. 
Multiple crossbars are connected in parallel to form the large-scale convolutional layer because of the size limitation of the memristor crossbar~\cite{7920854}. 
The ReLU activation function of CNN can be implemented by the integrate-and-fire circuits (IFCs) and other digital logic~\cite{7167197, 7920854}.
The deployment of deconvolutional network is similar to the convolutional layer. 
One major difference is the data should be zero-padded before giving to the crossbar, as is shown in Figure~\ref{pic:cnn_crossbar} (b). 
To optimally decrease the executions involved by the zeros, we transform and  group the input vectors with the zeros in the same locations, and only non-zero inputs in the rows are given to the crossbar and computed.

\section{DESIGN METHODOLOGY}
\label{sec:method}

\begin{figure}[b]
\centering
\includegraphics[width=0.45\textwidth]{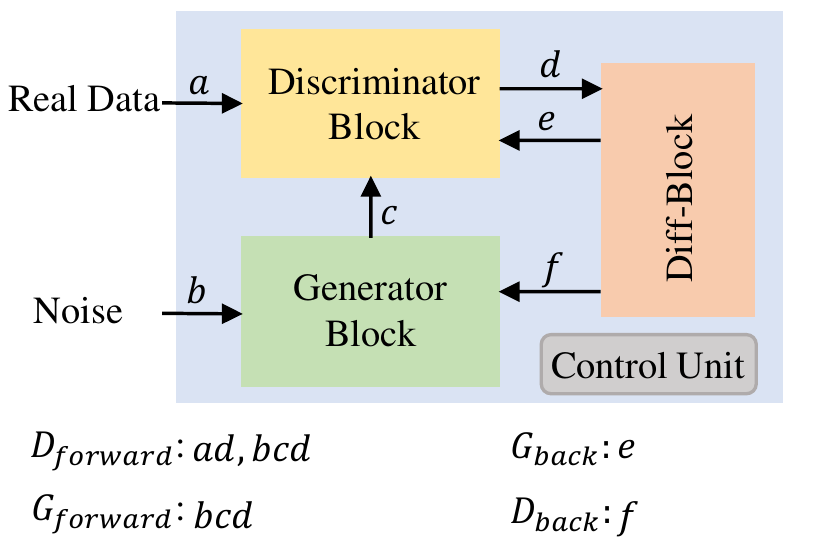}
\caption{Architecture of the proposed memristor-based GAN computation.}
\label{fig:architecture}
\end{figure}

In this section, we built a memristor-based neuromorphic design for accelerating both training and testing phase of GAN.
The basic computing architecture and data flow is described in subsection~\ref{sec:method:architecture}
To achieve optimal system performance, we proposed a cross-parallel pipline execution flow which is introduced in~\ref{sec:method:dataflow}.

\subsection{Memristor-based GAN Architecture }
\label{sec:method:architecture}

Based on the GAN computation described in~\ref{sec:GAN}, we developed the memristor-based computing architecture of GAN computation. 
As is shown in Figure~\ref{fig:architecture}, the proposed architecture is composed by four integrated components: Generator block, Discriminator block, Diff block, and Control unit. 
The functionality of these components can be summarized as below.

(1) The Discriminator block is designed to compute the $D(x^{i}$ and $D(G(z^{i})$ in Equation~\ref{equ:cost function1} for the stochastic gradient calculation.
It  is composed of a sea of connected memristor-based CNN units;

(2) The Generator block is built to calculate the $G(z^{i})$ in generating the artificial samples from the noise inputs for computing the stochastic gradient in generator weight updating.
Similarly, it  is composed of a sea of connected memristor-based DeCNN units;	

(3)Following Equation~\ref{equ:cost function1} and Equation~\ref{equ:cost function2}, the Diff block computes the gradients of the discriminator and generator block respectively.
The Diff block is constructed by the memristor-based circuits blocks including LUT (look up table), memory, adder, etc.

(4) The Control unit is designed to control the data flow and is built with combinational and sequential digital logic. 

In Figure~\ref{fig:architecture}, the basic data flow is demonstrated as $a\sim f$. 
The explanation of each step is summarized as below, where the definitions of $D(\cdot)$, $G(\cdot)$, $x$, $z$, and $m$ are as the same as those in subsection~\ref{sec:GAN}. 

\begin{itemize}
	\item$a$: The real data with $m$ samples in a batch, i.e. ${x^{i}}$ is given to the Discriminator block to compute the $D(x^i)$, where $i\in \{1,...,m\}$. 
	\item$b$: The noise data with $m$ samples in a batch, i.e. ${z^{i}}$ is given to the Generator block to obtain artificial samples $G(z^i)$, where $i\in \{1,...,m\}$.
	\item$c$: The $G(z^i)$ is generated from the Generator block and then transmitted to the Discriminator block to compute the $D(G(z^i))$, where $i\in \{1,...,m\}$.
	\item$d$: This step computes the transmits the $D(x^i)$ and $D(G(z^i))$ from the Discriminator block to the Diff block.  
	
	The above steps $a\sim d$ fulfill the forward computations of GAN.
	
	\item$e$: Based on the gradient calculated by the Diff block, i.e. $Error_{D}$ in Equation~\ref{equ:cost function1}, this step updates the weights of the Discriminator.
	\vspace{-4pt}
	\item$f$: The weights of the generator is updated in this step  according to the gradient obtained from the Diff block.
	
	These two steps \emph{e} and \emph{f} implement the backward computations in GAN.
\end{itemize}

Correspondingly, the $D_{forward}$ is composed of $a, c$ and $b, c, d$, $D_{back}$ is fulfilled by $e$, $G_{forward}$ is implemented by $b, c, d$, and $G_{back}$ is the data flow of $f$.



\begin{figure}[b]
\centering
\includegraphics[width=0.45\textwidth]{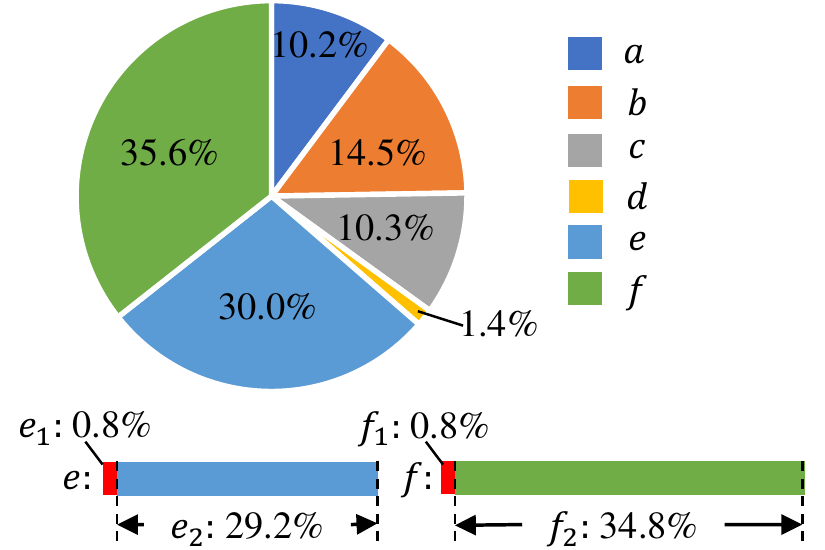}
\caption{Time cost of each step of the basic data flow.}
\label{fig:computing}
\end{figure}

\subsection{The Cross-Parallel Pipeline}
\label{sec:method:dataflow}

\begin{figure*}[!t]
\centering
\includegraphics[width=0.9\textwidth]{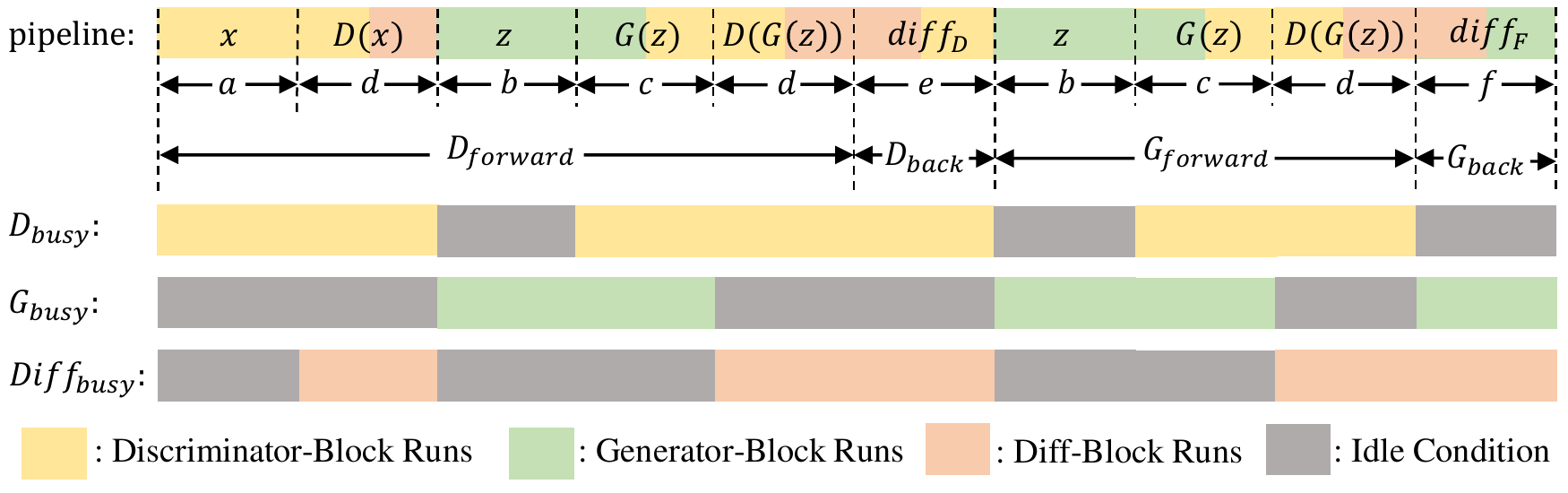}
\caption{The pipeline of each computing block based on the basic data flow.}
\label{fig:pipeline}
\end{figure*}

In this section, we proposed a cross-parallel pipeline based on the basic data flow for optimized computing efficiency. 
Hence, we first analyze the time cost of each step among $a\sim f$ in Section~\ref{sec:method:architecture}. 
The analysis is executed based on the simulations on NVSim simulator~\cite{NVSim} and the results are demonstrated in Figure~\ref{fig:computing}. 
In the analysis, the steps of \emph{e} and \emph{f} are divide into two categories: data transmission and data computation.
For example, $e_{1}$ refers to the gradients transmission time from the Diff block to the Discriminator block while $e_{2}$ represents the time consumed by updating the weights of the discriminator. 
Similarly, the $f_{1}$ is the gradients transmission and $f_{2}$ is the cost of weights updating of the generator. 
It is observed that the majority of the time cost occurs in updating the generator and discriminator. 

\begin{figure}[!b]
\centering
\includegraphics[width=0.45\textwidth]{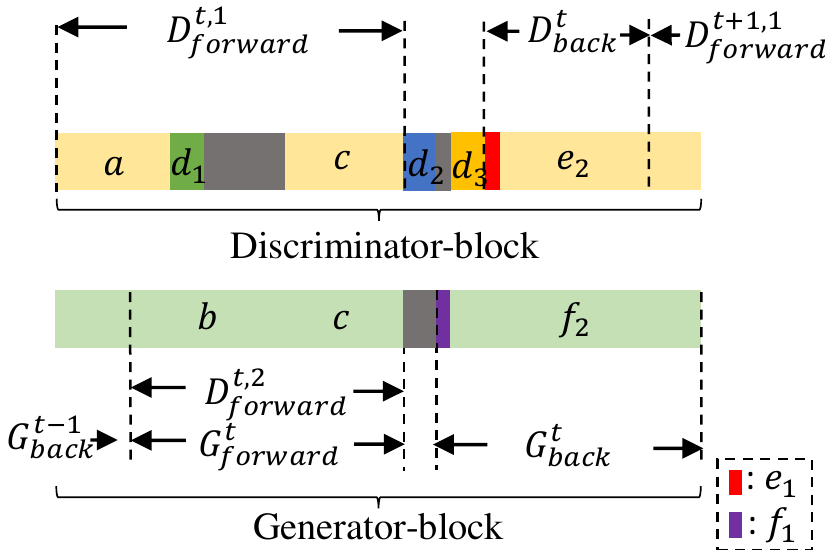}
\caption{The cross-parallel pipeline.}
\label{fig:cross}
\end{figure}

Figure~\ref{fig:pipeline} shows the pipline of each computing block, i.e. the Discriminator block, Generator block, and Diff block based on the basic data flow.
As is described, different colors represent the execution states of each block.
Then, we analyze the independence property each execution in these blocks to build the cross-parallel pipline, where the independent executions are designed to compute in parallel.
For example, observed from Figure~\ref{fig:pipeline}, the discriminator block and the generator block are independent from each other when the $a$ and $e_{2}$ are executing by the discriminator block or $b$ and $f_{2}$ are executing by the generator block. 
Hence, these computations can be optimized to be parallelism data flow.
There are only two conditions that the generator and discriminator block cannot work in parallel: first, during $b$ and $c$ executions; second, the weights updating has not finished. 
Otherwise, the generator and discriminator can work highly in parallel and the optimized cross-parallel pipline is depicted   in Figure~\ref{fig:cross}.


In the developed cross-parallel pipline, the $D_{forward}$ is divided into two parts--$D^{t,1}_{forward}$ and $D^{t,2}_{forward}$, where $t$ represents the $t^{th}$ iteration in GAN training. 
$D^{t,1}_{forward}$ includes the basic \emph{a} and \emph{d} execution and the $D^{t,2}_{forward}$ refers to the basic $b\sim d$ processing in the discriminator block.
In the cross-parallel pipline, the \emph{a} (i.e. real samples processing) and the \emph{b} (i.e. artificial samples generation) are executed simultaneously. 
It is observed from Figure~\ref{fig:computing} that \emph{b} is more time-consuming than \emph{a}, hence the $D(x^{i})$ that computed from \emph{a} is transmitted and stored in the Diff block firstly which is represented by $d_1$ in Figure~\ref{fig:cross}.
And a memory unit based on memristor crossbar is designed in the Diff block to store the computing results of $D(x^i)$. 
Then, the discriminator is idle which means no execution occurs until the artificial samples generation (i.e. \emph{b}) is finished. 
The $D(G(z))$ is transmitted to Diff block after the \emph{c} finished, which is represented by $d_2$ in Figure~\ref{fig:cross}. 
Immediately, $d_{3}$ which includes the gradient computation and its transmission to the discriminator block is executed.
Consequently, the weights updating in the discriminator ($e_2$) and generator ($f_2$) block run simultaneously.
In addition, although the $e_2$ is executed faster than $f_2$, the next training iteration of the discriminator starts asynchronously without introducing additional memory usage in the discriminator block as is indicated by $D^{t+1,1}_{forward}$. 
The training process of the generator and discriminator become synchronous before \emph{c} starts in the $(t+1)^{th}$ iteration.

\begin{table}[t]
\centering
\caption{Computation time of Basic Pipline vs. Cross-Parallel Pipeline}
\label{table-cross}
\begin{tabular}{|c|c|c|}
\hline
 & Basic Pipline & Cross-Parallel \\
\hline
Each Iteration(s) & 0.18& 0.11\\
$D_{Idle Time}(s)$ &0.069&0.018\\
$G_{Idle Time}(s)$ &0.033&0.015\\
\hline
\end{tabular}
\end{table}

As is discussed above, the parallelism of the computation can be largely improved. 
The time cost of the GAN computation following the basic pipline and the proposed cross-parallel pipline is evaluated on CIFAR-10 dataset~\cite{cifar10} to indicate this improvement.
Also, the simulation is executed on the simulator NVSim~\cite{NVSim}.
The results are shown in Table~\ref{table-cross}, we can observe that $1.6\times$ speedup is achieved by utilizing the cross-parallel pipline. 
In addition, the usage rates of the Discriminator block and Generator block are improved $3.8\times$ and $2.2\times$ respectively. 


\section{Design Detail}
\label{sec:detail}

In this section, the implementation of the Dicriminator block, Generator block and Diff block is explained in detail. 
All of the above blocks are implemented based on the memristor crossbars.

\subsection{Parallel and Memory-Free Structure for Discriminator and Generator Blocks}
In previous research~\cite{7920854}, the memristor-based computations for CNN training and testing has been implemented. 
However, these designs cannot be used in the GAN training because two major reasons.
First, the gradient for the output layer of CNN was computed as the output minus the true label previously, which is not adapted to GAN.
Moreover, the previous framework involves a large number of memory units, such a design will result in heavy area cost and time-consuming in GAN computations.

\begin{figure}[!t] 
\centering
\includegraphics[width=0.42\textwidth]{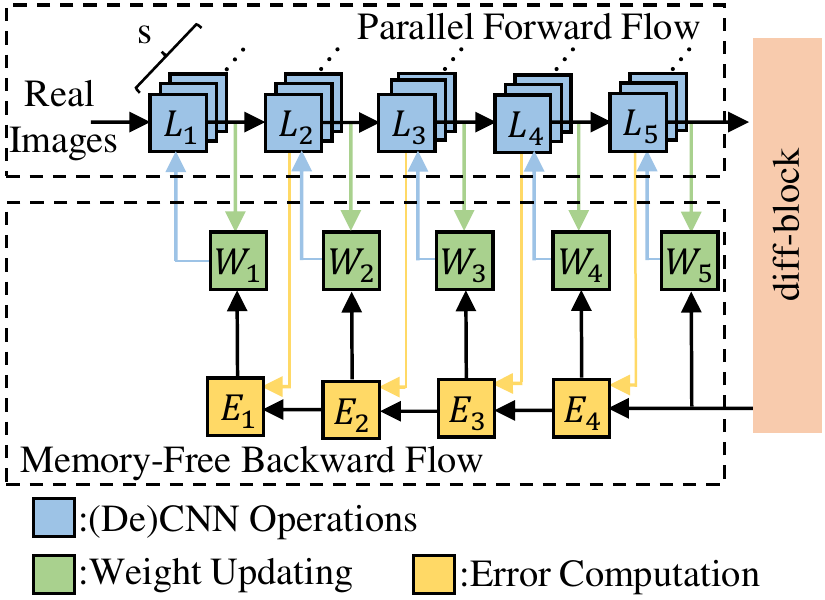}
\caption{The parallel and memory-free structure for the discriminator and generator blocks. The black arrows represent data input and output, while the colorful arrows represent the programming.}
\label{fig:structure}
\end{figure}

To solve the above challenges, we proposed a a parallel and memory-free structure as is shown in Figure~\ref{fig:structure}. The squares in different color represent the processing units of error computations, weight updating, and (de)convolutional operations respectively. These units are built by the memristor-based crossbar, IFC, and digital logic~\cite{7167197, 7920854}, and a simplified mapping scheme is shown in Figure~\ref{fig:map}. The proposed structure is composed of a parallel forward flow and a memory-free backward flow. It can work as the discriminator or generator block with different initial weights programmed on the (de)convolutional operation units. 


\subsubsection{The Parallel Forward Flow}

\begin{figure}[t] 
\centering
\includegraphics[width=0.45\textwidth]{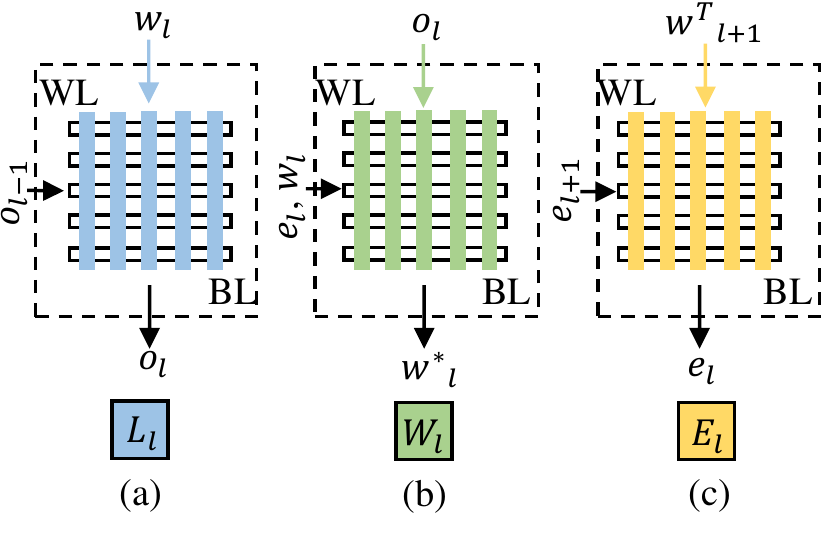}
\caption{The simplified mapping method of different layers. Blue: (de)convolutional operations, Green: weight updating, Yellow: error computation.}
\label{fig:map}
\end{figure}

The forward processing of GAN includes the CNN and DeCNN computations.
Consider a GAN structure with 5-layer CNN and 5-layer DeCNN, the parallel flow is depicted in the above part of Figure~\ref{fig:structure}.  
Initial or updated weights, i.e. $w_{l}$ in Figure~\ref{fig:map} (a) are programmed to the memristor crossbar of the (de)convolutional layer and the results processed from the forward flow are obtained as $o_{l}$ that works as the input to the next layer.
The mapping and programming method is follows Section~\ref{sec:crossbar} and previous researches~\cite{7920854,7167197}.
Multiple samples can be processed on these same processing units with different inputs, and thus parallel computations can be achieve.
The final output of the parallel forward flow is transmitted to the Diff block.



\subsubsection{The Memory-Free Backward Flow}

The memory-free backward flow aims to update the weights of the forward flow based on the gradient from the diff-block.

The backpropogation method used to update weights of (de)cnn can be summarized as equation~\ref{equ:backpropotion}, where $l$ represents the index of the (de)convolutional layer, $e_{l}$ represents the gradient, $w_{l}$ represents the weights, $w^{*}_{l}$ represents the updated weights and $w^{T}_{l}$ represents the transposition matrix of the weights, $o'_{l}$ represents the dirivative of the $l^{th}$ layer output signals and $\alpha$ represents the learning rate. Because the dirivative of ReLU activiation function is equal to the output signal itself, $o'_{l}$ is equal to $o_{l}$.

\begin{equation}\label{equ:backpropotion}
\left\{\begin{matrix}
e_{l}=e_{l+1}\bigotimes w^T_{l} & (a)\\ 
w^{*}_{l}=\alpha \cdot e_{l} \bigotimes o'_{l}+w_l&(b) 
\end{matrix}\right.  
\end{equation}

The memory-free backward flow is proposed to implement the backpropogation method, which is depicted as the below part of figure~\ref{fig:structure}. Weights' transposition, i.e. $w^{T}_{l}$ in Figure~\ref{fig:map} (c) are read from the operation uints in the forward flow and programmed to the memristor crossbar of the error computation uints. Outputs of each layer in the parallel forward flow, i.e. $o_{l}$ in Figure~\ref{fig:map} (b) are programmed to the memristor crossbar of the weight updating uints. The input to the error computation uint is the errors computed from the error commputation uint in the next layer, i.e. $e_{l+1}$ in Figure~\ref{fig:map} (c) and then current layers' errors, i.e. $e_{l}$ in Figure~\ref{fig:map} (c) are computed. These errors are also input to the weight updating uints to compute the updated weights i.e. $w^{*}_{l}$ in Figure~\ref{fig:map} (b).


In previous designs~\cite{7920854,7167197}, they use memory to store the updated weights or the inter-layer signals of CNN. The reason that the proposed design does not need special memory is that we store the weights and inter-layer signals in the same memristor crossbar as computation uints. The inter-layer signals are programmed to the weight updating uints directly used to compute the updated weights, shown as the green arrows in figure~\ref{fig:structure}. The updated weights are programmed to the (de)cnn computation uints, shown as the blue arrows in figure~\ref{fig:structure}.   

\subsubsection{Timing Sequence of the Parallel and Memory-Free Structure}
\begin{figure}[t] 
\centering
\includegraphics[width=0.45\textwidth]{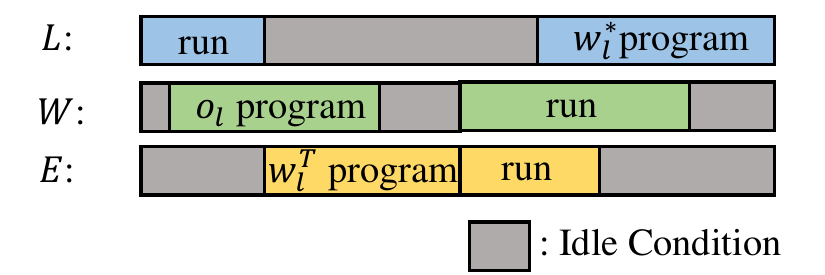}
\caption{The timing sequence of the parallel and memory-free structure}
\label{fig:timing}
\end{figure}

The timing sequence of one iteration for the parallel and memory-free structure is detailed as Figure~\ref{fig:timing}. First, (de)cnn operation uints process the input data and the inter-layer signal is programmed to the weight updating uints immediately. After (de)cnn operation uints processing the input data, the weights from these uints are programmed to the error computation uints and then they begin computing the backpropogation errors and weight updating uints begin computing the new weights which are programed to the (de)cnn operation uints directly.      

\subsection{The Implementation of Diff-Block}
\begin{figure}[b]
\centering
\includegraphics[width=0.4\textwidth]{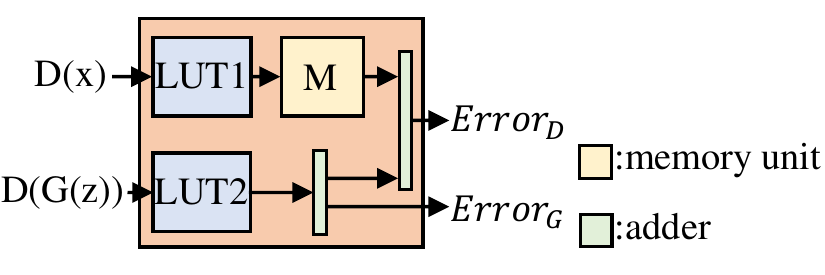}
\caption{The architecture of diff-block}
\label{pic:diff-block}
\end{figure}

The implementation of Diff-Block is detailed shown as Figure~\ref{pic:diff-block}. Diff-block is composed by two memristor-based look-up tables (LUT)~\cite{ReRAMLUT}, the memristor-based memory unit \cite{NVSim,Niu2012DesignTF} and two memristor-based adders~\cite{7167197}. LUT1 stores values of $\triangledown _{\theta} logD(x)$ and LUT2 stores values of $\triangledown _{\theta} log(1-D(G(z)))$. Linear transfromation, $\frac{1}{m}$, can be done in the adder~\cite{7167197}.   
   
When the discriminator-block transmitted $D(x)$ to the diff-block, values of $\triangledown _{\theta} logD(x)$ are read from the LUT1 and stored in the memristor-based memory. The memory should be able to store $m$ values where $m$ represents the batch size. Generally, the batch size is 64. When the discriminator-block transmitted $D(G(z))$ to the diff-block, values of $\triangledown _{\theta} log(1-D(G(z)))$ are read from LUT2. Those values are input to the adder and the $Error_{G}$ is computed. Meantime, the data in the memory are input to the adder as well as the above result and $Error_{D}$ is computed. 

\section{EVALUATION}
In this section, we evaluate the performance of the proposed memristor-based GAN accelerator on accuracy, speed, energy, and area cost.
The performance is compared with the previous GPU-based platform~\cite{DCGAN} and the microarchtural design~\cite{fpgagan} based on FPGA. 
The Nvidia Geforce GTX 1080 is used as the GPU platform.
The proposed design is evaluated on NVSim~\cite{NVSim} simulator environment.
The memristor crossbar size is designed to be $32\times 32$, the resistance range of the memristor device is set to be $[50K\Omega, 1M\Omega]$, and the required crossbar number is calculated following the implementation in~\ref{sec:crossbar} and~\cite{PRIME}.
The circuits designs for the neurons and control units follows the~\cite{7920854}.

\subsection{Accuracy in Different Data Precision}
\begin{figure}[t]
\centering
\includegraphics[width=0.4\textwidth]{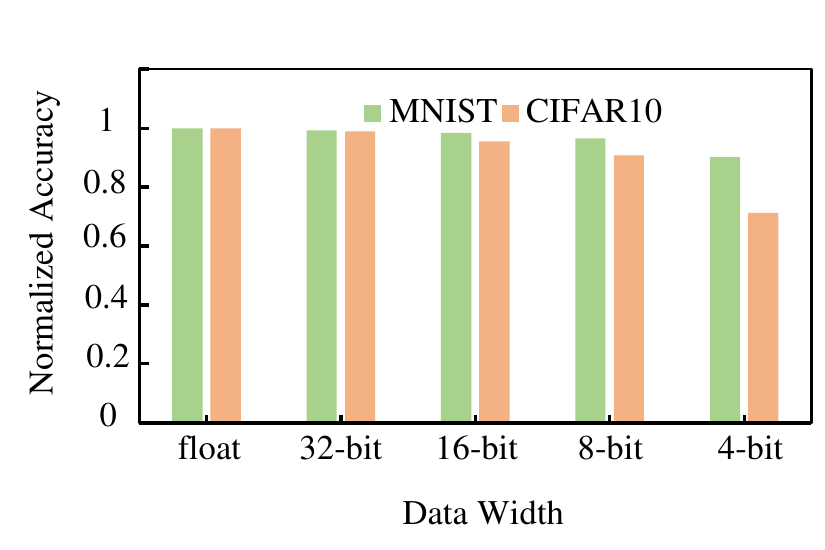}
\caption{Accuracy in different data precision.}
\label{pic:resolution}
\end{figure}

In general, the memristor supports limited precision in data transmission, data storage and computation~\cite{6033439,7167197}. 
Although analog states of memristor have been reported by the HP Research Lab~\cite{HP}, high precision involves in scarification of speed and design cost.
In this section, the data bitwidth for optimized accuracy and design cost is explored, and the GAN computing accuracy in different data bitwidth is shown in Figure~\ref{pic:resolution}.
As referred in \cite{DCGAN}, the performance of GAN can be measured by using the discriminator as a feature extractor for a classifier. In this experiment, SVM works as the classifer and the discriminator of GAN works as the extractor for image classification. MNIST and Cifar-10, whose characters are listed as table~\ref{table-datasets}, are used as the benchmark. The accuracy based on the feature extractor trained from GPU is regarded as the baseline. The accuracy based on the fixed-point feature extractors is compared with the baseline.  
Two dataset is utilized as is depicted in Table~\ref{table-datasets}: MNIST and CIFAR-10.

\begin{table}[!b]
\centering
\caption{Dataset}
\label{table-datasets}
\begin{tabular}{|c|c|c|c|c|}
\hline
 & training size & testing size & scale & category\\
\hline
MNIST & 60000&10000&$20 \times 20$& 10 \\
CIFAR-10 &50000&10000&$32 \times 32$& 10\\
\hline
\end{tabular}
\end{table}

Figure~\ref{pic:resolution} shows the normalized accuracy when discriminators are trained by data in different data format and bitwidth. 
The results show that the system accuracy with 8-bit data precision has a slight accuracy loss the the normalized accuracy is still higher than 90\% on MNIST and CIFAR10.
However, significant accuracy loss is introduced in the 4-bit data precision.
Therefore, the 8-bit data precision, i.e. memristor device with 8-bit states is utilized in this work and the following evaluation.

\subsection{Computing Parallelism}

In the training process of GAN, the generator processes a batch of noise samples and the discriminator processes a batch of real images.
The computing parallelism is referred to the generator (or discriminator) blocks that compute in parallel in each training iteration, i.e. the \emph{s} in Figure~\ref{fig:structure} 
As is indicated in Section~\ref{sec:method:dataflow}, the computing efficiency can be improved by higher computing parallelism.
Note that higher parallelism results in an increase of the area and design cost in the proposed GAN computing system.
In this section, the speed and area cost in different data parallelism scenarios are explored.
The bedroom dataset from Lsun~\cite{yu15lsun} is used in this evaluation. 
The maximum parallelism size is selected to be 64.


\begin{figure}[t]
\centering
\includegraphics[width=0.4\textwidth]{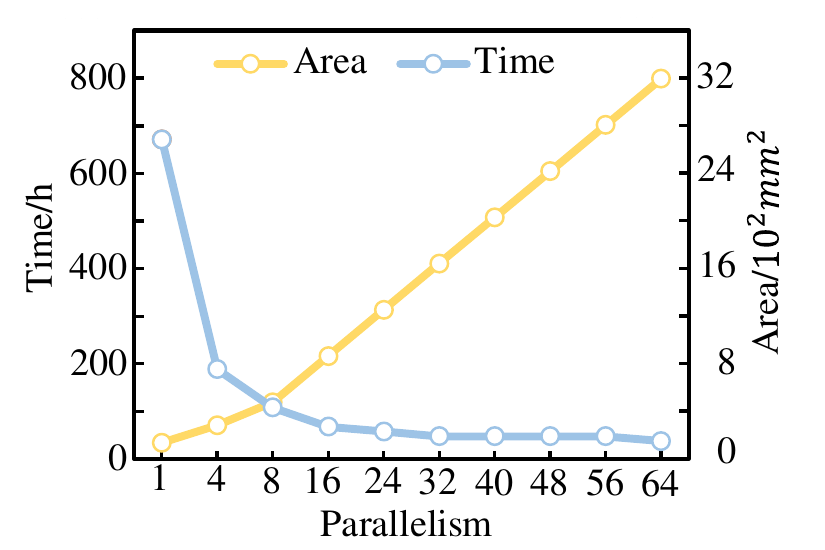}
\caption{Time and area cost on different computing parallelism}
\label{pic:parallelism}
\end{figure}

Figure~\ref{pic:parallelism} shows the time and area cost in different parallelism. 
The computation time decreases with the increase of the parallelism size while the area cost increases heavily in high parallelism scenario.
The main reason is that parallel computing can accelerate the speed of the image generation in the generator block, as well as the real and artificial images generation in the discriminator block.
However, the speed of other procedures in training such as propagation and weight updating in the backward computations does not rely on the parallelism improvement. 
Hence, the speed increase rate becomes extremely slow in large parallelism designs while the area cost increases fast.
Based on the results, the computing parallelism size is set to be 32 in the developed memristor-based GAN accelerator.


\subsection{Memristor-based GAN Computing Speed}
\label{subsection:efficiency of the ReRAM-Based Accelerator}

\begin{table}[b]
\centering
\caption{Dataset}
\label{table-big-datasets}
\begin{tabular}{|c|c|c|}
\hline
 & Dateset Size & Image Size \\
\hline
ImageNet & 456567& $400 \times 300$ \\
Lsun/bedroom &3033042& $256 \times 256$\\
\hline
\end{tabular}
\end{table}

In this section, the speed of the memristor-based GAN accelerator is evaluated and compared with conventional platforms such as GPU and FPGA in previous works.
Two big dataset is chosen--ImageNet and Lsun/bedroom is chosen for better demonstration and the dataset detail is listed in Table~\ref{table-big-datasets}.
As is discussed above, the data bitwidth in the proposed memristor-based design is 8 and the parallelism size is 32.

The batch size is set to be 64 in these three scenarios.
The speed of GAN training process is evaluated based on the fact that the testing process also works as an inter-step in GAN training.

\begin{table}[t]
\centering 
\caption{Speed of the memristor-based, GPU, FPGA-based accelerator}
\label{table-efficiency-comparison}
\begin{tabular}{|c|c|c|c|c|}
\hline
\multirow{2}{*}{} & \multicolumn{2}{|c|}{ImageNet} & \multicolumn{2}{|c|}{LSUN/bedroom}  \\
\cline{2-5} &time(h)&speedup&time(h)&speedup\\
\hline
This Work & 6.3&-&47.2&- \\
GPU & 17 &$2.7\times$&130&$2.8\times$\\
FPGA & 30 &$4.8\times$&255&$5.5\times$\\
\hline
\end{tabular}
\end{table}

The experimental results are listed in Table~\ref{table-efficiency-comparison}
It is observe that our proposed design can achieve $2.8\times$ ($2.7\times$) compared with GPU and $5.5\times$ ($4.8\times$) speedup compared with the FPGA-based accelerator on Lsun (ImageNet).
In addition, our design has higher speedup for larger dataset because the parallel pipeline and the memory-free structure  can largely decrease the time cost as is discussed in Section~\ref{sec:detail}.


\begin{figure}[t]
\centering 
\subfigure[Proposed Design]
{\label{fig:efficiency:a} \includegraphics[width=0.2\textwidth]{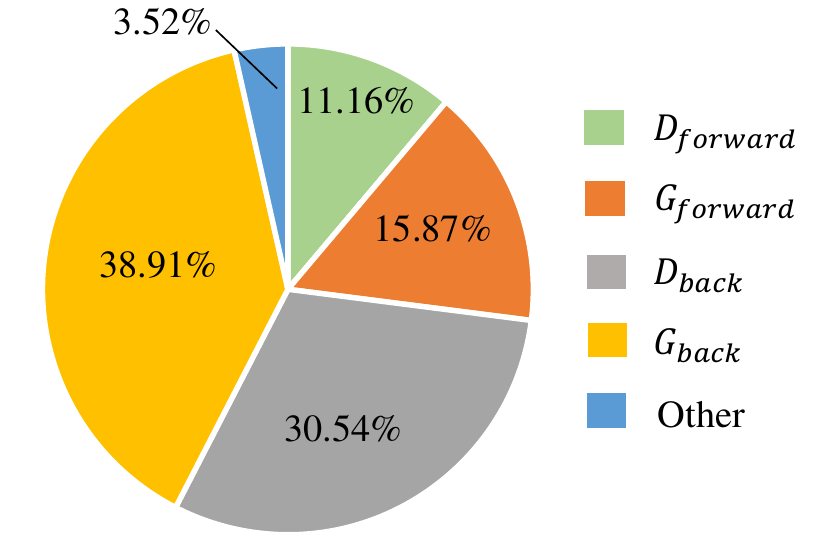}} 
\subfigure[GPU]
{\label{fig:efficiency:b} \includegraphics[width=0.2\textwidth]{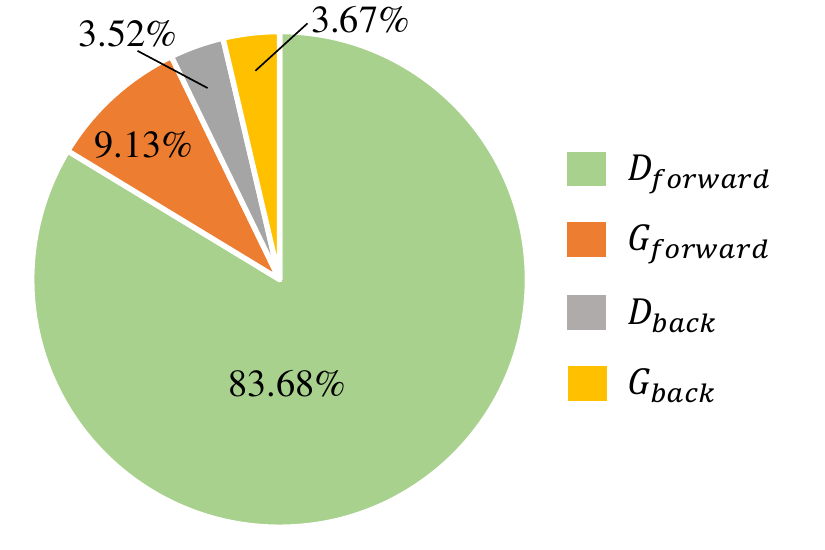}} 
\caption{Time cost for each procedure of training GAN} 
\label{fig:efficiency_procedure} 
\end{figure}

We also analyze the time cost of each computing procedures of GAN training that described in Section~\ref{sec:GAN}, and the results are indicated in Figure~\ref{fig:efficiency_procedure}. 
As is shown in Figure~\ref{fig:efficiency_procedure} (b), major time cost in the GPU based GAN computations occurs in the $D_{forward}$ procedure.
However, our design reduces such a cost efficiently from 83.68\% to 11.16\%, as is demonstrated in Figure~\ref{fig:efficiency_procedure} (a).
Our proposed design performs highly resource usage compared with the GPU own to the developed cross-parallel pipline, hence the GAN computing speed is improved efficiently.



\subsection{Energy and Area Cost Analysis}

In this section, we evaluate and compare the energy and area cost of the memristor-based GAN accelerator. 
The energy cost of each procedure in GAN training is also analyzed in detail. 

\begin{table}[b]
\centering 
\caption{Energy Cost of the ReRAM/GPU/FPGA-based accelerator}
\label{table-energy-comparison}
\begin{tabular}{|c|c|c|c|c|}
\hline
\multirow{3}{*}{} & \multicolumn{2}{|c|}{ImageNet} & \multicolumn{2}{|c|}{LSUN/bedroom}  \\
\cline{2-5} &energy&\multirow{2}{*}{saving} &energy&\multirow{2}{*}{saving}\\
&($KW/h$)&&($KW/h$)&\\
\hline
This Work& 0.51&-&3.8&- \\
GPU & 3.1&$6.1\times$&23.4&$6.1\times$\\
FPGA & 0.79&$1.5\times$&5.5&$1.4\times$\\
\hline
\end{tabular}
\end{table} 

Table~\ref{table-energy-comparison} shows the energy cost comparison of the memristor-, GPU-, and FPGA-based accelerator. 
Our proposed design achieves $6.1\times$ and $1.4\times$ energy saving compared with the GPU and FPGA-based GAN computing respectively. 
The energy cost of each training procedure is analyzed as Figure~\ref{fig:efficiency_procedure}. 
It is observed that the energy cost of the $D_{forward}$ in the proposed design has a lower energy cost percentage in the whole training process compared with the computations on GPU.
Such a low energy cost owns to our proposed memory free data flow in the memristor-based accelerator, in which the energy cost of data communication between memory and CNN (or DeNN) is saved.


The area of the proposed design is $1644mm^2$ when the parallelism size is designed to be 32. 
The parallel forward flows in the discriminator-block and generator-block accounts for $44.8\%$ and $49.7\%$ respectively. The area of the parallel forward flow is positively related to the parallelism size. 
When the parallelism size is equal to 1, the area of parallel flows in discriminator and generator-block are equal to $23.0mm^2$ and $25.5mm^2$ respectively and the total area is $139mm^2$.


\begin{figure}[t]
\centering 
\subfigure[Proposed Design]
{\label{fig:energy:a} \includegraphics[width=0.2\textwidth]{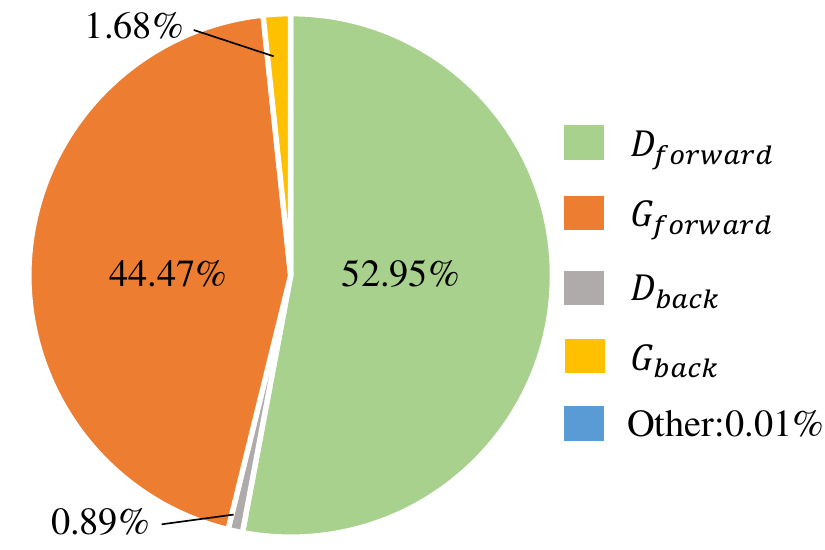}} 
\subfigure[GPU]
{\label{fig:energy:b} \includegraphics[width=0.2\textwidth]{efficiency_subprocedure_gpu.pdf}} 
\caption{Energy cost for each procedure of training GAN} 
\label{fig:energy_procedure} 
\end{figure}

\section{CONCLUSION}
Generative adversarial network (GAN) is an effective unsupervised model that is extremely computationally expensive. To address this issue, we proposed a memristor-based accelerator. The proposed design has two major aspects including a cross-parallel pipeline and the memory-free flow. The proposed accelerator was tested on large dataset: ImageNet and Lsun. With area equal to $1644mm^2$, the proposed accelerator can achieve $2.8\times$ ($2.7\times$) and $5.5\times$ ($4.8\times$) speedup, as well as $6.1\times$ ($6.1\times$) and $1.4\times$ ($1.5\times$) energy-saving compared with GPU-based and FPGA-based GAN accelerators respectively on Lsun (ImageNet) dataset.


\bibliographystyle{ieeetr}
\small
\bibliography{sample-bibliography.bib} 

\end{document}